\documentclass{mem}
\usepackage{natbib}
\usepackage{txfonts}
\usepackage{balance}
\usepackage{graphicx}
\usepackage{flushend}

\usepackage{xcolor}
\usepackage[caption=false]{subfig}

\usepackage[breaklinks,pdftex]{hyperref}
\idline{75}{282}
\begin{document}
\def\teff{$T\rm_{eff }$}
\def\kms{$\mathrm {km s}^{-1}$}

\title{
Towards a TeV blazar sequence and its physical interpretation
}

   \subtitle{}

\author{
I. \,Viale\inst{1,2} 
\and E. \,Prandini\inst{1,2} \and C. \,Righi\inst{3} \and F. \,Bovolon\inst{1} \and N. \,Sahakyan\inst{4}
          }

\institute{
Dipartimento di Fisica e Astronomia, Università degli Studi di Padova, I-35131 Padova, Italy\\
\email{ilaria.viale@unipd.it}
\and
INFN, Sezione di Padova, I-35131 Padova, Italy
\and
INAF, Osservatorio Astronomico di Brera, I-23807 Merate, Italy
\and 
ICRANet-Armenia, Marshall Baghramian Avenue 24a, Yerevan 0019, Armenia
}

\authorrunning{Viale}

\titlerunning{TeV blazar sequence}

\date{Received: XX-XX-XXXX (Day-Month-Year); Accepted: XX-XX-XXXX (Day-Month-Year)}

\abstract{
Blazars, a highly energetic subclass of jetted active galactic nuclei, show a broad band spectral energy distribution (SED) with two bumps, resulting from non-thermal jet emission. In 1998, an anticorrelation between the SED luminosity and the peaks frequency was found, later confirmed in 2017, called the blazar sequence. Its origin is still unclear.
This work is part of a broader effort aiming at giving a physical interpretation to the blazar sequence, by modeling the sources emission in a Synchrotron Self Compton framework, concentrating only on TeV-detected blazars of BL Lac type. Unlike the original sequence, sources were binned by synchrotron peak frequency.  The SED of one representative source is modeled for each bin, using data from average activity state to ensure consistency.
Here, we focus on the study of one of the selected representative sources, PKS~2155-304, a high-energy-peaked BL Lac, showing the performed data selection and preliminary modeling results.



}
\maketitle{}

\section{Introduction}
Blazars are a subclass of active galactic nuclei (AGN) dominating the extragalactic $\gamma-$ray sky.
They host a relativistic jet of particles pointing 
in the direction of 
Earth.
This jet is responsible for the broad-band  emission typically observed from these objects, extending from radio to $\gamma-$rays, reaching TeV energies in some cases. 
The Spectral Energy Distribution (SED) of blazars is characterized by two bumps, ascribed to the non-thermal processes undergone by the accelerated particles populating the jet.

One of the first comprehensive studies on 
blazars' spectral properties
have been performed by \citet{Fossati1998}. 
Comprising sources from X-ray and radio samples and dividing them into radio-luminosity bins, the authors found an anticorrelation between the bolometric luminosity of the emitting sources and the frequency of their SED peaks, a trend called the \emph{blazar sequence}.
In addition, also an increase of the Compton Dominance, i.e. the ratio between peaks luminosity (higher-energy peak luminosity divided by the lower-energy one), was observed.
This trend was later confirmed in 2017, with a larger dataset comprising \emph{Fermi-}LAT \citep{fermi_Atwood2009} sources divided into $\gamma-$ray luminosity bins \citep{sequence2.0}.
It is more prominent when considering both blazar classes, i.e. Flat Spectrum Radio Quasars (FSRQ) and BL Lac objects, with FSRQs mostly populating the highest luminosity bins, but remains still present also when considering BL Lacs only, which span 5 orders of magnitude in the low-energy peak frequency (from $\sim 10^{17}$ Hz to $\sim10^{12}$ Hz), when increasing luminosity by 4 orders of magnitude.

In recent years, attempts were made to include TeV sources in the sequence.
Among them, \citet{PrandiniGhisellini2022} applied the same analysis as \citet{sequence2.0} to the discovered TeV blazars.
They found no significant differences in the average SEDs compared to previous works; except for an increased X-ray luminosity for the brightest TeV BL Lacs. 

In these proceedings, we present a selection of TeV-detected blazars to be investigated in the context of the blazar sequence.
For consistency, only BL Lacs were considered.
Then we focus on the selected high-energy-peaked BL Lac (HBL) PKS~2155-304, providing an interpretation of its emission using the simplest model proposed in the literature: the Synchrotron Self Compton model \citep[SSC,][]{ssc_marashi1992}. 
This study serves as an initial step toward understanding the effect of these objects on the blazar sequence.





%
%

\section{Source selection}

\begin{figure}
\resizebox{\hsize}{!}{\includegraphics[clip=true]{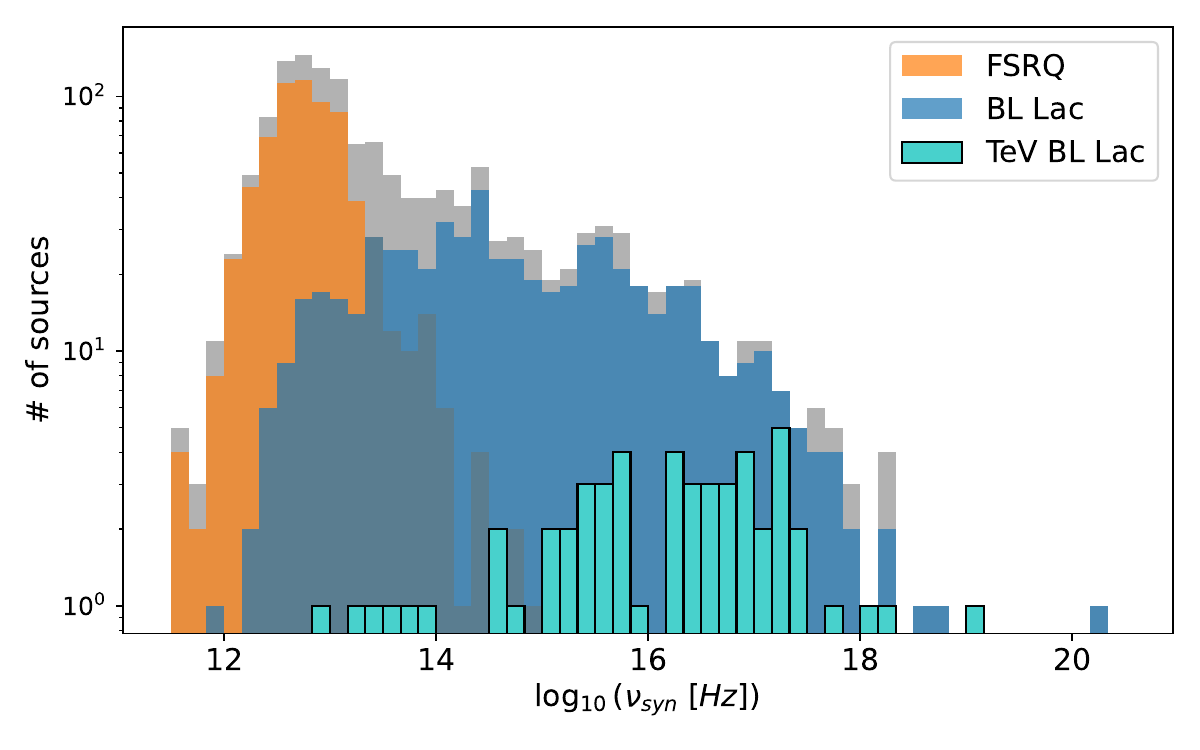}}
\caption{
\footnotesize
Synchrotron peak frequency distribution of 4LAC-DR2 blazars. All sources are shown in grey, FSRQs in orange, BL Lacs in blue, TeV-detected BL Lacs in cyan.}
\label{fig:nusyn_distr_4lac}
\end{figure}
\begin{figure*}[t!]
\centering
\resizebox{0.9\hsize}{!}{\includegraphics[clip=true]{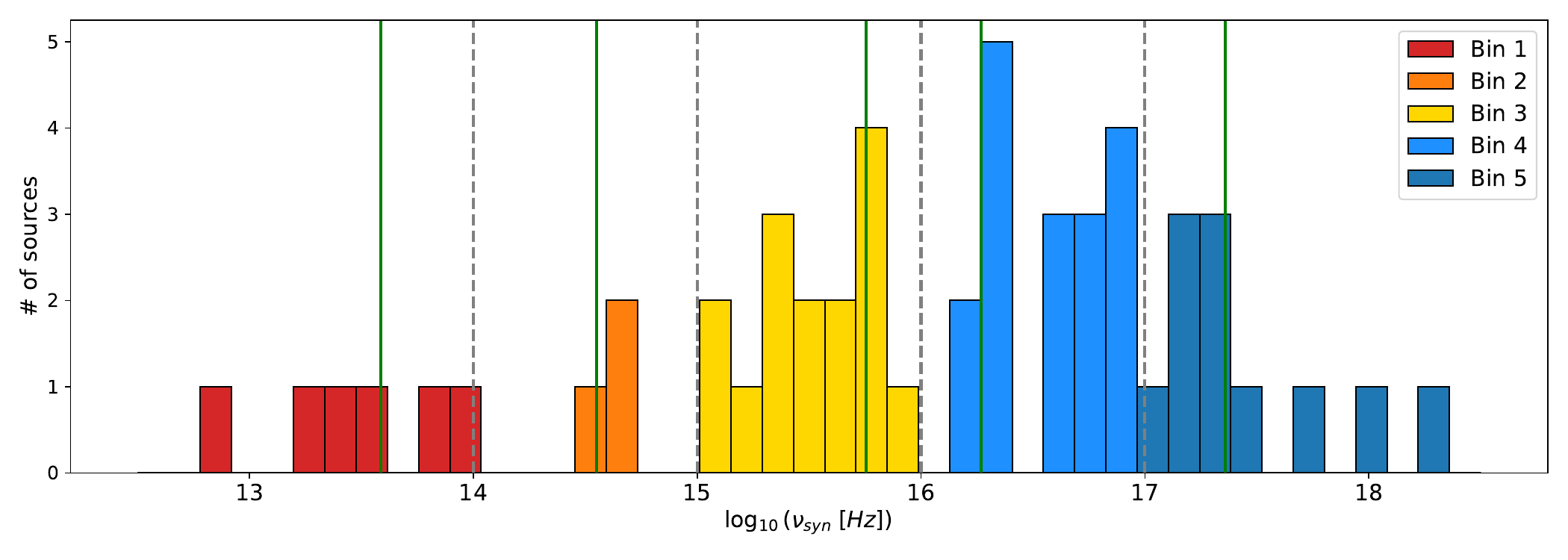}}
\caption{\footnotesize
{Synchrotron peak frequency-based distribution and bins defined for our study. Only TeV BL Lac objects are shown. The vertical green lines indicate the $\nu_{syn}$ of the selected representative sources.
}
}
\label{fig:nusyn_distr_tev-only}
\end{figure*}
\begin{table*}[ht!]
\small
\label{ttab:representative-sources}
\begin{center}
\begin{tabular}{lccc}
\hline
 & Bin [Hz] & Source & $\nu_{syn}$ [Hz] \\
\hline
Bin 1 & $\nu_{syn} < 10^{14}$ & BL Lac & $3.86\times10^{13}$   \\
Bin 2 & $10^{14} < \nu_{syn} < 10^{15}$ & TXS~0506+056 & $3.55\times10^{14}$ \\
Bin 3 & $10^{15} < \nu_{syn} < 10^{16}$ & PKS~2155-304 & $5.69\times10^{15}$ \\
Bin 4 & $10^{16} < \nu_{syn} < 10^{17}$ & PG~1218+304 & $1.85\times10^{16}$ \\
Bin 5 & $\nu_{syn} > 10^{17}$ & PKS~0548-322 & $2.29\times10^{17}$ \\
\hline
\end{tabular}
\end{center}
\caption{\footnotesize Representative sources chosen for each bin. Columns show: bin number, frequency range, source name, synchrotron peak frequency.}
\end{table*}

To date, only 84 blazars have been detected at Very High Energies (VHE, $E>100$ GeV)\footnote{http://tevcat2.uchicago.edu/}, of which 68 are BL Lacs.
We selected information about these sources from the second release of the Fourth Catalog of AGNs \citep[4LAC-DR2][]{4LAC, 4LAC-DR2}, based on the first 10 years of \emph{Fermi-}LAT observations. 
We concentrated only on BL Lacs with known redshift, needed to produce a precise modeling of the sources emission, getting a final sample of 55 objects.

In previous works on the blazar sequence, sources were divided into bins based on their luminosity in a pre-defined band.
Then, the single-blazar SEDs in each bin were 
averaged in order to get the average emission representing the bin.
While well established, this approach has risks, since
a) it biases the selection to the luminosity in a single band, and b) combines the emission of sources with possibly different spectral features and variability.

In this work we face the blazar sequence study with a different approach.
In order not to be biased by the luminosity in a single band, we divided the sources into bins based on their synchrotron peak frequency, $\nu_{syn}$.
Fig.~\ref{fig:nusyn_distr_4lac} shows the $\nu_{syn}$ distribution of TeV detected BL Lacs together with that of all 4LAC-DR2 blazars, while Fig.~\ref{fig:nusyn_distr_tev-only} provides a zoom in the TeV-detected BL Lac distribution, showing the division into frequency bins (by decade) and the number of objects for each bin.
Regarding the sources emission in the bins, we decided not to create an average SED in order not to lose the spectral properties of the single sources, but instead select a representative source per bin.
The main arguments driving the choice were its location in the bin (not too close to the bin edges) and the presence of a generous amount of VHE $\gamma-$ray data.
The chosen sources are listed in Table~\ref{ttab:representative-sources}.

\section{Data selection}

\begin{figure}
\resizebox{\hsize}{!}{\includegraphics[clip=true]{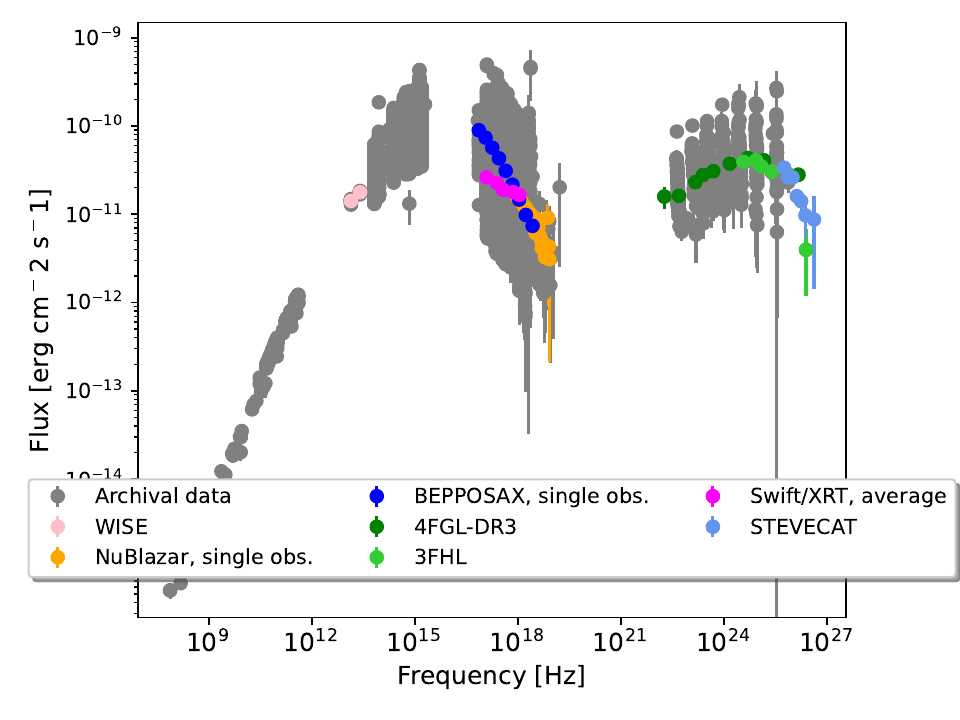}}
\caption{
\footnotesize
Data selection for the source PKS 2155-304. representative of the third bin ($10^{15} < \nu_{syn} < 10^{16}$).}
\label{fig:bin3_data-selection}
\end{figure}

For each of the five representative sources, the SED was built.
Multi-wavelength data were taken from the Markarian Multiwavelength Data Center database\footnote{https://mmdc.am/}\citep[MMDC,][]{mmdc_Sahakyan2024}, 
while
VHE $\gamma-$ray data were taken from from the Spectral TeV Extragalactic Catalog \citep[STeVECat,][]{stevecat_Greaux2023}.
In order to build the SED of each source, data corresponding to an average state of activity were selected. 
However, 
since it was not always possible to rely on simultaneous multi-wavelength data for the selected objects, we concentrated the data selection on the sources' activity state, gathering data from different data-taking periods.
In addition, given the different duty cycle of the instruments involved and the high variability of the objects in the X-ray band, we followed two different approaches: a) we averaged the X-ray data, as it is often done in these kind of studies, in order to obtain the average emission of the source; and b) we selected single X-ray observations corresponding to an average activity state, since it is possible that with a simple average the results may be biased towards the activity states when most of the observations were performed. The two selections can be found in Fig~\ref{fig:bin3_data-selection} for the source PKS~2155-304, representative of the third bin ($10^{15} < \nu_{syn} < 10^{16}$).
Note the high variability of this source in the X-ray band, as underlined by the archival data, and that the two selections are quite different, hence affecting the final modeling results presented in Sect.~\ref{sec:modeling}.



\section{Modeling} \label{sec:modeling}

The recovered SEDs were modeled within the framework of an SSC model \citep{ssc_marashi1992}, the simplest leptonic model available for the description of blazars emission.
As a starting point, we concentrated on high-energy-peaked BL Lacs (HBL), namely sources with $\nu_{syn} > 10^{15}$ Hz, since objects with lower $\nu_{syn}$ are difficult to describe with a pure SSC model,
as they host denser sources of external photon fields in their environment, making the effect of the External Compton process more prominent and necessary to well interpret their emission.
Therefore, in the following, we will model only the representative BL Lacs from the three higher-frequency bins and compare their results.

To perform the source modeling, we used MMDC \citep{mmdc_Sahakyan2024}, an online modeling software based on Convolutional Neural Netowrks (CNN) \cite{cnn-model_Begue2024}.
The CNN governing the software was trained by the authors on 
simulated blazar spectra, generated with the SOPRANO code \citep{soprano_Gasparyan2022}.
Thus, MMDC allows to perform self-consistent time-dependent blazar modeling.
The advantage resides in the tool Deep Learning nature, as it allows to significantly reduce the modeling computational time and scan the whole parameter space of the model, avoiding it to fall in a local minimum.

\begin{figure*}
\centering
\subfloat[]{%
  \includegraphics[width=0.45\textwidth]{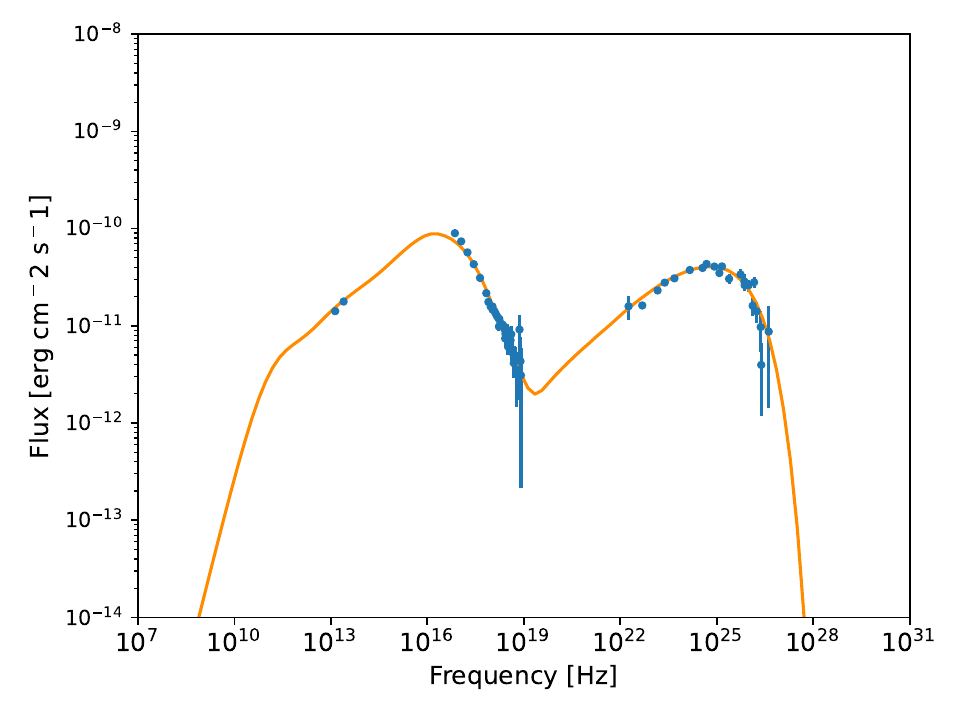
  }%
  \label{fig:sed_pks2155_singleXrayobs}%
}\qquad
\subfloat[]{%
  \includegraphics[width=0.45\textwidth]{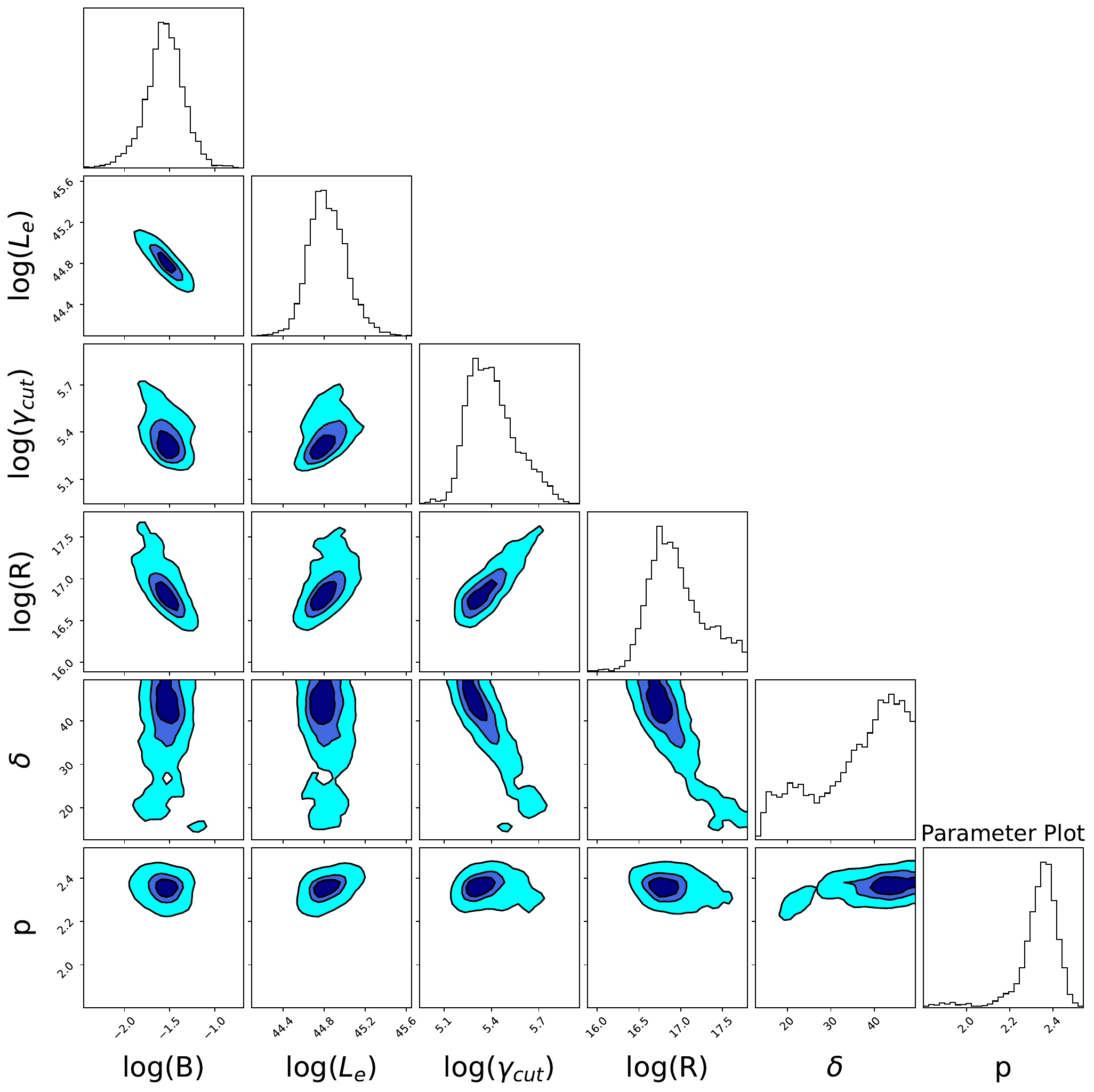}%
  \label{fig:corner-plot_pks2155_singleXrayobs}%
}\\
\caption{SSC model (left) and related corner plot (right) of the source PKS~2155-304. In this SED the data selection comprises single X-ray observations.}
\label{fig:model_pks2155}
\end{figure*}

MMDC results are reported in Table \ref{tab:modeling_results} for the source PKS~2155-304, taking into account both the selections with the single X-ray observations and the X-ray average.
Fig.~\ref{fig:model_pks2155} shows the model obtained for the same source with the single X-ray observations in the SED data, together with its corner plot describing the
relationship between the free parameters investigated in the model.



\begin{table*}[]
\label{tab:modeling_results}
\begin{center}
\begin{tabular}{l|c|c}
\hline
& \multicolumn{2}{c}{PKS~2155-304} \\
& \multicolumn{1}{c}{X-ray average} & \multicolumn{1}{|c}{Single X-ray obs.}  \\
\hline
t$_{var}$ {[}$\times10^4$ s{]} & $48.8$ & $4.86$  \\
p1 & 2.24 & 2.34 \\
log(L$_e$ {[}erg/s{]}) & 44.9 & 44.77 \\
$\log(\gamma_{max})$ & 5.45 & 5.28 \\
$\delta$ & 21.47 & 43.06 \\
log(B {[}G{]}) & 1.70 & -1.51 \\ 
\hline
\end{tabular}
\end{center}
\caption{\footnotesize Modeling parameters obtained for both data selections of the source PKS~2155-304. 
} 
\end{table*}

\section{Discussion and conclusions}


In this contribution, we
performed an initial step toward the investigation of
the role of TeV sources in the blazar sequence, 
with the final aim of
giving a physical interpretation of the sequence itself through a self-consistent leptonic modeling of the emission of the selected objects.
In order to investigate the blazar sequence, we divided the TeV blazars into $\nu_{syn}$ bins and selected one representative candidate per bin, whose emission was then modeled.
Here, we present the results from on the source PKS~2155-304, comparing the modeling obtained with the data selections described above, and planning to model a larger number of sources in order to find trends hinting at the physical mechanisms underlying the sequence.

The data selection resulted to be complex, especially in X-rays, were the selected objects show a very high variability. For this reason, two approaches were employed: one based on the average of X-ray data, and the other on single X-ray observations.
Given the different slope of X-ray data with these selections (see Fig.~\ref{fig:bin3_data-selection}), the results on the modeling of the same source can be quite different, as shown in Table \ref{tab:modeling_results}.
Here, the main differences rely on the values of the variability time scale, $\mathrm{t_{var}}$, and the Doppler factor, $\delta$.
Focusing on $\delta$, the large difference in the resulting best parameters can be ascribed to its poorly constrained distribution in the modeling (see Fig.~\ref{fig:corner-plot_pks2155_singleXrayobs}), which makes these results less reliable.
To solve this issue we started investigating models resulting from the assumption of different $\delta$ values, aiming at a better interpretation of the modeling results.

Since both data selections are reasonable and well motivated, it is not trivial to indicate the best results. A potential solution for the future could involve continuous of sources in X-rays, in order not to be biased by the specific state of activity of the source at the time of the observation; or to have online tools filtering sources observations on the period, in order to select simultaneous data.

\begin{acknowledgements}
IV and EP acknowledge the project “SKYNET: Deep Learning for Astroparticle Physics”, PRIN 2022 (CUP: D53D23002610006).
\end{acknowledgements}

\small
\bibliographystyle{aa}
\bibliography{bibliography}

\begin{thebibliography}{11}
\expandafter\ifx\csname natexlab\endcsname\relax\def\natexlab#1{#1}\fi

\bibitem[{{Ajello} {et~al.}(2020){Ajello}, {Angioni}, {Axelsson}, {Ballet}, {Barbiellini}, {Bastieri}, {Becerra Gonzalez}, {Bellazzini}, {Bissaldi}, {Bloom}, {Bonino}, {Bottacini}, {Bruel}, {Buson}, {Cafardo}, {Cameron}, {Cavazzuti}, {Chen}, {Cheung}, {Ciprini}, {Costantin}, {Cutini}, {D'Ammando}, {de la Torre Luque}, {de Menezes}, {de Palma}, {Desai}, {Di Lalla}, {Di Venere}, {Dom{\'\i}nguez}, {Dirirsa}, {Ferrara}, {Finke}, {Franckowiak}, {Fukazawa}, {Funk}, {Fusco}, {Gargano}, {Garrappa}, {Gasparrini}, {Giglietto}, {Giordano}, {Giroletti}, {Green}, {Grenier}, {Guiriec}, {Harita}, {Hays}, {Horan}, {Itoh}, {J{\'o}hannesson}, {Kovac'evic'}, {Krauss}, {Kreter}, {Kuss}, {Larsson}, {Leto}, {Li}, {Liodakis}, {Longo}, {Loparco}, {Lott}, {Lovellette}, {Lubrano}, {Madejski}, {Maldera}, {Manfreda}, {Mart{\'\i}-Devesa}, {Massaro}, {Mazziotta}, {Mereu}, {Meyer}, {Migliori}, {Mirabal}, {Mizuno}, {Monzani}, {Morselli}, {Moskalenko}, {Negro}, {Nemmen}, {Nuss}, {Ojha}, {Ojha}, {Omodei}, {Orienti}, {Orlando}, {Ormes},
  {Paliya}, {Pei}, {Pe{\~n}a-Herazo}, {Persic}, {Pesce-Rollins}, {Petrov}, {Piron}, {Poon}, {Principe}, {Rain{\`o}}, {Rando}, {Rani}, {Razzano}, {Razzaque}, {Reimer}, {Reimer}, {Schinzel}, {Serini}, {Sgr{\`o}}, {Siskind}, {Spandre}, {Spinelli}, {Suson}, {Tachibana}, {Thompson}, {Torres}, {Torresi}, {Troja}, {Valverde}, {van Zyl}, \& {Yassine}}]{4LAC}
{Ajello}, M., {Angioni}, R., {Axelsson}, M., {et~al.} 2020, \apj, 892, 105

\bibitem[{{Atwood} {et~al.}(2009){Atwood}, {Abdo}, {Ackermann}, {Althouse}, {Anderson}, {Axelsson}, {Baldini}, {Ballet}, {Band}, {Barbiellini}, {Bartelt}, {Bastieri}, {Baughman}, {Bechtol}, {B{\'e}d{\'e}r{\`e}de}, {Bellardi}, {Bellazzini}, {Berenji}, {Bignami}, {Bisello}, {Bissaldi}, {Blandford}, {Bloom}, {Bogart}, {Bonamente}, {Bonnell}, {Borgland}, {Bouvier}, {Bregeon}, {Brez}, {Brigida}, {Bruel}, {Burnett}, {Busetto}, {Caliandro}, {Cameron}, {Caraveo}, {Carius}, {Carlson}, {Casandjian}, {Cavazzuti}, {Ceccanti}, {Cecchi}, {Charles}, {Chekhtman}, {Cheung}, {Chiang}, {Chipaux}, {Cillis}, {Ciprini}, {Claus}, {Cohen-Tanugi}, {Condamoor}, {Conrad}, {Corbet}, {Corucci}, {Costamante}, {Cutini}, {Davis}, {Decotigny}, {DeKlotz}, {Dermer}, {de Angelis}, {Digel}, {do Couto e Silva}, {Drell}, {Dubois}, {Dumora}, {Edmonds}, {Fabiani}, {Farnier}, {Favuzzi}, {Flath}, {Fleury}, {Focke}, {Funk}, {Fusco}, {Gargano}, {Gasparrini}, {Gehrels}, {Gentit}, {Germani}, {Giebels}, {Giglietto}, {Giommi}, {Giordano}, {Glanzman},
  {Godfrey}, {Grenier}, {Grondin}, {Grove}, {Guillemot}, {Guiriec}, {Haller}, {Harding}, {Hart}, {Hays}, {Healey}, {Hirayama}, {Hjalmarsdotter}, {Horn}, {Hughes}, {J{\'o}hannesson}, {Johansson}, {Johnson}, {Johnson}, {Johnson}, {Johnson}, {Kamae}, {Katagiri}, {Kataoka}, {Kavelaars}, {Kawai}, {Kelly}, {Kerr}, {Klamra}, {Kn{\"o}dlseder}, {Kocian}, {Komin}, {Kuehn}, {Kuss}, {Landriu}, {Latronico}, {Lee}, {Lee}, {Lemoine-Goumard}, {Lionetto}, {Longo}, {Loparco}, {Lott}, {Lovellette}, {Lubrano}, {Madejski}, {Makeev}, {Marangelli}, {Massai}, {Mazziotta}, {McEnery}, {Menon}, {Meurer}, {Michelson}, {Minuti}, {Mirizzi}, {Mitthumsiri}, {Mizuno}, {Moiseev}, {Monte}, {Monzani}, {Moretti}, {Morselli}, {Moskalenko}, {Murgia}, {Nakamori}, {Nishino}, {Nolan}, {Norris}, {Nuss}, {Ohno}, {Ohsugi}, {Omodei}, {Orlando}, {Ormes}, {Paccagnella}, {Paneque}, {Panetta}, {Parent}, {Pearce}, {Pepe}, {Perazzo}, {Pesce-Rollins}, {Picozza}, {Pieri}, {Pinchera}, {Piron}, {Porter}, {Poupard}, {Rain{\`o}}, {Rando}, {Rapposelli}, {Razzano},
  {Reimer}, {Reimer}, {Reposeur}, {Reyes}, {Ritz}, {Rochester}, {Rodriguez}, {Romani}, {Roth}, {Russell}, {Ryde}, {Sabatini}, {Sadrozinski}, {Sanchez}, {Sander}, {Sapozhnikov}, {Parkinson}, {Scargle}, {Schalk}, {Scolieri}, {Sgr{\`o}}, {Share}, {Shaw}, {Shimokawabe}, {Shrader}, {Sierpowska-Bartosik}, {Siskind}, {Smith}, {Smith}, {Spandre}, {Spinelli}, {Starck}, {Stephens}, {Strickman}, {Strong}, {Suson}, {Tajima}, {Takahashi}, {Takahashi}, {Tanaka}, {Tenze}, {Tether}, {Thayer}, {Thayer}, {Thompson}, {Tibaldo}, {Tibolla}, {Torres}, {Tosti}, {Tramacere}, {Turri}, {Usher}, {Vilchez}, {Vitale}, {Wang}, {Watters}, {Winer}, {Wood}, {Ylinen}, \& {Ziegler}}]{fermi_Atwood2009}
{Atwood}, W.~B., {Abdo}, A.~A., {Ackermann}, M., {et~al.} 2009, \apj, 697, 1071

\bibitem[{B{\ifmmode\acute{e}\else\'{e}\fi}gu{\ifmmode\acute{e}\else\'{e}\fi} {et~al.}(2024)B{\ifmmode\acute{e}\else\'{e}\fi}gu{\ifmmode\acute{e}\else\'{e}\fi}, Sahakyan, Dereli-B{\ifmmode\acute{e}\else\'{e}\fi}gu{\ifmmode\acute{e}\else\'{e}\fi}, Giommi, Gasparyan, Khachatryan, Casotto, \& Pe'er}]{cnn-model_Begue2024}
B{\ifmmode\acute{e}\else\'{e}\fi}gu{\ifmmode\acute{e}\else\'{e}\fi}, D., Sahakyan, N., Dereli-B{\ifmmode\acute{e}\else\'{e}\fi}gu{\ifmmode\acute{e}\else\'{e}\fi}, H., {et~al.} 2024, Astrophys. J., 963, 71

\bibitem[{Fossati {et~al.}(1998)Fossati, Maraschi, Celotti, Comastri, \& Ghisellini}]{Fossati1998}
Fossati, G., Maraschi, L., Celotti, A., Comastri, A., \& Ghisellini, G. 1998, Monthly Notices of the Royal Astronomical Society, 299, 433

\bibitem[{Gasparyan {et~al.}(2022)Gasparyan, B{\ifmmode\acute{e}\else\'{e}\fi}gu{\ifmmode\acute{e}\else\'{e}\fi}, \& Sahakyan}]{soprano_Gasparyan2022}
Gasparyan, S., B{\ifmmode\acute{e}\else\'{e}\fi}gu{\ifmmode\acute{e}\else\'{e}\fi}, D., \& Sahakyan, N. 2022, Mon. Not. R. Astron. Soc., 509, 2102

\bibitem[{{Ghisellini} {et~al.}(2017){Ghisellini}, {Righi}, {Costamante}, \& {Tavecchio}}]{sequence2.0}
{Ghisellini}, G., {Righi}, C., {Costamante}, L., \& {Tavecchio}, F. 2017, \mnras, 469, 255

\bibitem[{Gr{\ifmmode\acute{e}\else\'{e}\fi}aux {et~al.}(2023)Gr{\ifmmode\acute{e}\else\'{e}\fi}aux, Biteau, Hassan, Hervet, Nievas~Rosillo, \& Williams}]{stevecat_Greaux2023}
Gr{\ifmmode\acute{e}\else\'{e}\fi}aux, L., Biteau, J., Hassan, T., {et~al.} 2023, arXiv e-prints, arXiv:2304.00835

\bibitem[{{Lott} {et~al.}(2020){Lott}, {Gasparrini}, \& {Ciprini}}]{4LAC-DR2}
{Lott}, B., {Gasparrini}, D., \& {Ciprini}, S. 2020, arXiv e-prints, arXiv:2010.08406

\bibitem[{{Maraschi} {et~al.}(1992){Maraschi}, {Ghisellini}, \& {Celotti}}]{ssc_marashi1992}
{Maraschi}, L., {Ghisellini}, G., \& {Celotti}, A. 1992, \apjl, 397, L5

\bibitem[{Prandini \& Ghisellini(2022)}]{PrandiniGhisellini2022}
Prandini, E. \& Ghisellini, G. 2022, Galaxies, 10, 35

\bibitem[{Sahakyan {et~al.}(2024)Sahakyan, Vardanyan, Giommi, B{\ifmmode\acute{e}\else\'{e}\fi}gu{\ifmmode\acute{e}\else\'{e}\fi}, Israyelyan, Harutyunyan, Manvelyan, Khachatryan, Dereli-B{\ifmmode\acute{e}\else\'{e}\fi}gu{\ifmmode\acute{e}\else\'{e}\fi}, \& Gasparyan}]{mmdc_Sahakyan2024}
Sahakyan, N., Vardanyan, V., Giommi, P., {et~al.} 2024, Astron. J., 168, 289

\end{thebibliography}

\end{document}